\documentclass{article}

\usepackage[T1]{fontenc} 
\usepackage[utf8]{inputenc} 
\usepackage{ismir,amsmath}
\usepackage{url}
\usepackage{graphicx}
\usepackage{color}
\usepackage{multirow}
\usepackage{amsfonts}       
\usepackage[dvipsnames, table]{xcolor}
\usepackage{pifont} 

\usepackage[inline]{enumitem}
\usepackage[backend=biber,style=numeric-comp, maxnames=10, doi=false,isbn=false,url=false,eprint=false,date=year,sorting=none]{biblatex}
\AtEveryBibitem{%
  \clearfield{note}%
  \clearfield{editor}
    \clearname{editor}
    \clearfield{pages}
     \clearlist{location}
}
\definecolor{mygreen}{rgb}{0.032, 0.6392, 0.2039}
\newcommand{\xmark}{\ding{55}}%

\usepackage{booktabs,arydshln}
\makeatletter
\def\adl@drawiv#1#2#3{%
        \hskip.5\tabcolsep
        \xleaders#3{#2.5\@tempdimb #1{1}#2.5\@tempdimb}%
                #2\z@ plus1fil minus1fil\relax
        \hskip.5\tabcolsep}
\newcommand{\cdashlinelr}[1]{%
  \noalign{\vskip\aboverulesep
           \global\let\@dashdrawstore\adl@draw
           \global\let\adl@draw\adl@drawiv}
  \cdashline{#1}
  \noalign{\global\let\adl@draw\@dashdrawstore
           \vskip\belowrulesep}}
\makeatother

\makeatletter
\renewcommand{\paragraph}{%
  \@startsection{paragraph}{4}%
  {\z@}{0.8ex \@plus 1ex \@minus .2ex}{-1em}%
  {\normalfont\normalsize\bfseries}%
}
\makeatother

\def\bench {MuChoMusic}
\newcommand{\numQuestions}{1,187}
\newcommand{\numTracks}{644}

\title{MuChoMusic: Evaluating Music Understanding in Multimodal Audio-Language Models}

\multauthor
{Benno Weck*$^1$ \hspace{1cm} Ilaria Manco*$^2$ \hspace{1cm} Emmanouil Benetos$^2$} { \bfseries{Elio Quinton$^3$ \hspace{1cm} George Fazekas$^2$ \hspace{1cm} Dmitry Bogdanov$^1$}\\
$^1$Universitat Pompeu Fabra, $^2$Queen Mary University of London, $^3$Universal Music Group\\
{\small * equal contribution} 
{\tt\small benno.weck01@estudiant.upf.edu, i.manco@qmul.ac.uk}
}

\def\authorname{B. Weck, I. Manco, E. Benetos, E. Quinton, G. Fazekas, and D. Bogdanov}

\begin{document}
\maketitle
\begin{abstract}
Multimodal models that jointly process audio and language hold great promise in audio understanding and are increasingly being adopted in the music domain. By allowing users to query via text and obtain information about a given audio input, these models have the potential to enable a variety of music understanding tasks via language-based interfaces. However, their evaluation poses considerable challenges, and it remains unclear how to effectively assess their ability to correctly interpret music-related inputs with current methods. Motivated by this, we introduce MuChoMusic, a benchmark for evaluating music understanding in multimodal language models focused on audio. MuChoMusic comprises \numQuestions{} multiple-choice questions, all validated by human annotators, on \numTracks{} music tracks sourced from two publicly available music datasets, and covering a wide variety of genres. Questions in the benchmark are crafted to assess knowledge and reasoning abilities across several dimensions that cover fundamental musical concepts and their relation to cultural and functional contexts. Through the holistic analysis afforded by the benchmark, we evaluate five open-source models and identify several pitfalls, including an over-reliance on the language modality, pointing to a need for better multimodal integration. Data and code are open-sourced.\footnote{Data: \url{https://doi.org/10.5281/zenodo.12709974}, website: \url{https://mulab-mir.github.io/muchomusic}}
\end{abstract}

\begin{figure}[t]
\centering
 \includegraphics[width=0.89\columnwidth]{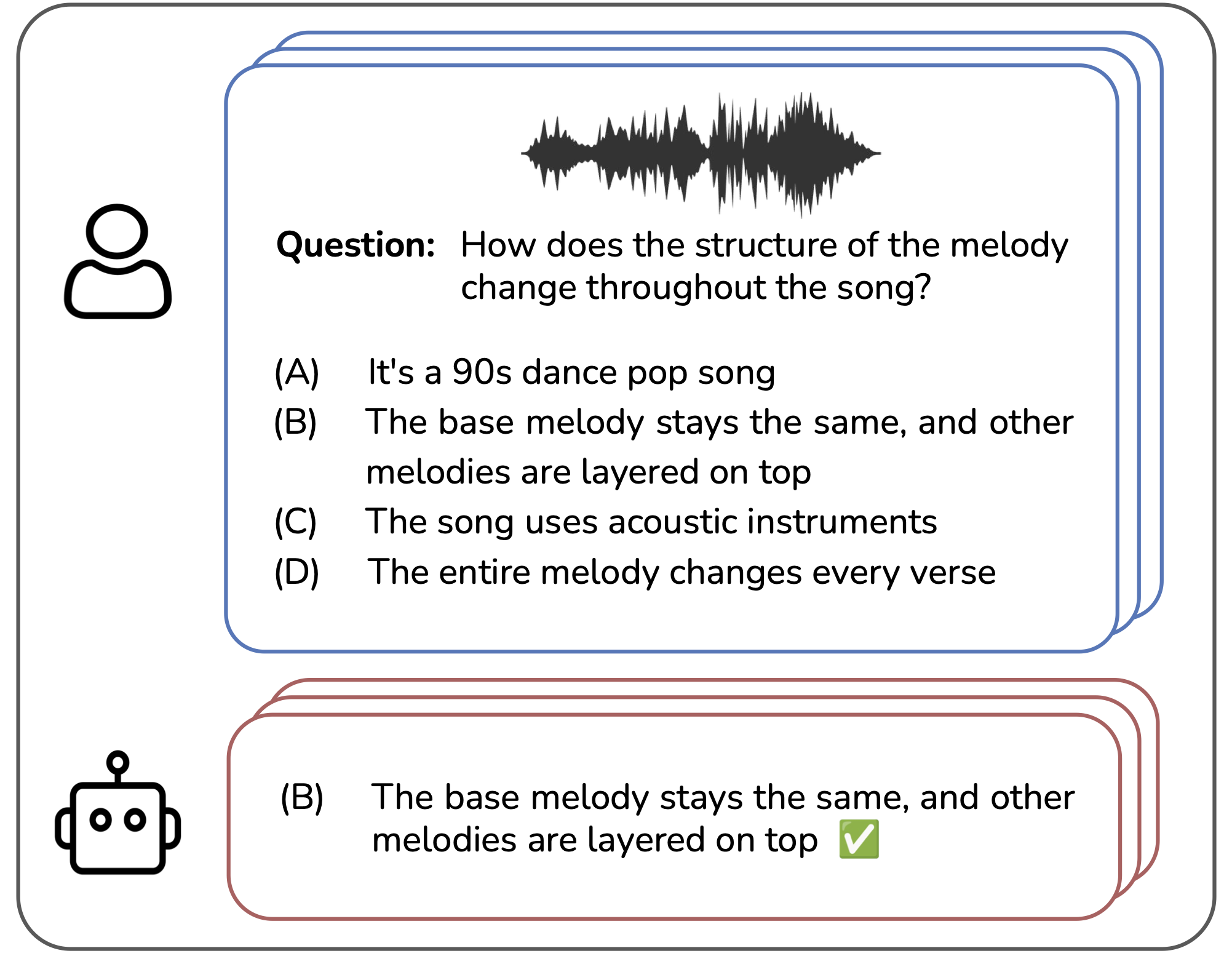}
  \vspace{-0.6em}
 \caption{\textbf{Multiple-choice questions} in \bench{} have four answer options of different levels of difficulty.}
 \label{fig:overview}
  \vspace{-0.6em}
\end{figure}

\section{Introduction}\label{sec:introduction}
Combining the success of large language models (LLMs) with new advances in machine perception that have led to image, audio and video foundation models \cite{bommasani_opportunities_2022}, multimodal LLMs are becoming influential across many fields \cite{peng2024grounding,gemini_team_gemini_2023, openai_gpt-4_2024, alayrac_flamingo_2022, bai_qwen-vl_2023}. Recently, models of this kind have started supporting the audio modality, with a subset also being applied to the music domain \cite{gardner_llark_2023, liu_music_2024, deng_musilingo_2023, deshmukh_pengi_2024, tang_salmonn_2024, hussain_m2ugen_2024, chu_qwen-audio_2023}. We refer to such models exhibiting audio understanding capabilities as \textit{Audio LLMs}. In a nutshell, Audio LLMs consist of pre-trained LLMs whose input space has been expanded beyond text to include tokens from an audio encoder, granting them the ability to produce language outputs that require understanding of both modalities. While promising, these models also inherit many of the limitations of LLMs and little attention has so far been given to their evaluation. In most cases, current automatic evaluation relies on match-based metrics which measure the semantic or lexical overlap between model outputs and reference text. However, many works have pointed out deficiencies in this approach \cite{bitton_visit-bench_2023}, which fails to capture the large space of acceptable language outputs admitted by open-ended tasks. For example, the question \textit{``What are some possible uses for this music in a film or TV show?''} may be suitably answered in many different ways. Secondly, automatic music understanding evaluation via language is only supported by a handful of human-annotated datasets \cite{agostinelli_musiclm_2023, manco_song_2023, mckee_language-guided_2023}, of which only one \cite{agostinelli_musiclm_2023} has widely been adopted in the context of Audio LLMs. Instead, many prior works have created a variety of ad-hoc datasets built upon synthetically generated captions from tags and other metadata \cite{doh_lp-musiccaps_2023, liu_music_2024, deng_musilingo_2023} to train and evaluate their models, without explicit data validation mechanisms, which raises questions around their reliability. These three key issues, lack of standardisation, the inadequacy of  text generation metrics, and the quality of annotations in current datasets, pose obstacles to the development of the field and has prompted some to resort to human evaluations \cite{gardner_llark_2023}, which can be costly and are hard to scale and reproduce. 

In this paper, we present \textit{\bench{}}, the first benchmark for evaluating music understanding in Audio LLMs. We design a test that is easy to evaluate by collecting a set 
of multiple-choice (MC) questions that are scrutinised by human annotators, on which simple classification accuracy can be obtained as a reliable indicator of music understanding over the categories covered by the test. The content of our benchmark is intended to be challenging, grounded in factual music knowledge, and tests core understanding and reasoning skills across several dimensions such as music theory, musical styles and traditions, historical and social contexts, structure and expressive analysis. Using \bench{}, we carry out a comprehensive evaluation of five existing Audio LLMs with music understanding capabilities. 
We envision that \bench{} will complement prior efforts to standardise music understanding evaluation \cite{bittner_mirdata_2019, yuan_marble_2023, plachouras_mir_ref_2023} by including this new family of models and steering their early development towards robust progress.

\section{Related Work}\label{sec:related_work}
In the music domain, Audio LLMs are commonly evaluated by assessing their text output in the context of a given task defined by an instruction template. Tasks are either designed to test whether the model is able to recognise predefined musical properties such as key (\textit{``What is the key of this song?''}), genre, instrumentation, etc., or they probe for outputs that encompass a variety of musical concepts and that more closely resemble the dialogue format typical of chatbots. Tasks that fall under the former usually mirror canonical MIR tasks and their evaluation leverages standard metrics and benchmarks from the MIR literature. Evaluation of tasks that require broader understanding follows instead less established protocols. Prior works on Audio LLMs most commonly tackle this via two tasks, music captioning (\textit{``Describe the contents of the provided audio in detail''}) \cite{tang_salmonn_2024, deng_musilingo_2023, liu_music_2024, gardner_llark_2023} and music question answering (\textit{``What are some possible uses for this music in a film or TV show?''}) \cite{liu_music_2024, deng_musilingo_2023}. To perform this kind of evaluation, the authors in \cite{tang_salmonn_2024, gardner_llark_2023, deng_musilingo_2023} make use of the MusicCaps dataset \cite{agostinelli_musiclm_2023}, while others \cite{liu_music_2024, deng_musilingo_2023} employ ad-hoc evaluation datasets created with the help of LLMs. In particular, Liu et al. \cite{liu_music_2024} and Deng et al. \cite{deng_musilingo_2023} propose their own datasets for music question answering, MusicQA and MusicInstruct respectively. These are derived from captions in the MusicCaps dataset or tags from the MagnaTagaTune dataset \cite{Law2009} (MusicQA only), by augmenting them into music-question pairs via pre-trained LLMs. Similarly to these works, we also leverage LLMs to generate our set of questions and answers, but we follow a multiple-choice format to ensure meaningful evaluation and validate all generated data through human annotators to guarantee high data quality.

Finally, we note that concurrent work also proposes evaluation benchmarks for music understanding in LLM-based models \cite{wang_muchin_2024, yuan_chatmusician_2024, li_music_2024}, but these all differ from our work in significant ways: MuChin \cite{wang_muchin_2024} includes only text in Chinese and does not follow a multiple-choice format, while both MusicTheoryBench (MTB) and ZIQI-Eval focus on the symbolic domain and address the evaluation of text-based LLMs. AIR-Bench \cite{yang_air-bench_2024} includes a small subset of music-related tasks, but puts its focus on audio understanding more generally.
We provide an overview of key differences with other benchmarks in Table \ref{tab:related_work}. 

\begin{table}[t]
\footnotesize
\centering
\begin{tabular}{p{1.955cm}cp{1.955cm}cp{0.3cm}>{\centering\arraybackslash}m{0.275cm}>{\centering\arraybackslash}m{0.3cm}}
\toprule
Benchmark &  Size   & Source(s) & Audio & HC  & MC   \\ 
\midrule
MusicQA \cite{liu_music_2024} & 4.5k & MagnaTagATune & \checkmark & \xmark & \xmark  \\
MusicInstruct \cite{deng_musilingo_2023}  & 61k & MusicCaps & \checkmark & \xmark & \xmark  \\
ZIQI-Eval \cite{li_music_2024} & 14k & Music books & \xmark & \xmark & \checkmark \\
MTB \cite{yuan_chatmusician_2024} & 372 & (human-written) & \xmark & \checkmark &  \checkmark \\
AIR-Bench \cite{yang_air-bench_2024} & 400 & MusicCaps & \checkmark & \xmark & \checkmark \\
MuChin \cite{wang_muchin_2024} & 1k & \textit{unknown} & \checkmark & \checkmark & \xmark  \\
\cdashlinelr{1-6}
\bench{} & 1.2k & MusicCaps, SDD & \checkmark & \checkmark & \checkmark \\
\bottomrule
\end{tabular}
\caption{\textbf{Comparison of \bench{} to existing benchmarks.} HC: human-curated, MC: multiple-choice.}
\label{tab:related_work}
 \vspace{-1em}
\end{table}

\section{MuChoMusic}\label{sec:muchomusic}
Through \bench{}, we aim to alleviate three prominent issues in the evaluation of music understanding in Audio LLMs: a lack of standardisation, the inadequacy of existing text generation metrics, and the quality of current evaluation sets.
We address the first two by adopting a multiple-choice format, while our methodical generation and validation procedure attends to the third issue by grounding the data in human-written descriptions and ensuring that the final questions and answers are correct and contextually relevant, as judged by multiple annotators.\footnote{We provide a datasheet \cite{gebru_datasheets_2021} with details about the dataset content, collection and validation in the online supplementary materials.}

\subsection{Overview}\label{sec:overview}
\bench{} consists of \numQuestions{} multiple-choice questions aimed at testing the understanding of \numTracks{} unique music tracks sourced from the MusicCaps \cite{agostinelli_musiclm_2023} and the Song Describer Dataset \cite{manco_song_2023}.
We adopt a multiple-choice format in order to standardise evaluation and  follow widespread practice in LLM-centric evaluation scenarios \cite{hendrycks_measuring_2020, zhong_agieval_2023, liang2023holistic, srivastava2023beyond}.
As illustrated in Figure \ref{fig:overview}, each question has four possible answers. One option is the correct answer, the other three are distractors. Inspired by \cite{berzak_starc_2020}, we structure these as follows: one does not fit the track of interest but is related to the question (\textit{incorrect but related}), one correctly fits the audio, but does not address the question (\textit{correct but unrelated}), and one does not apply to the track and is also irrelevant to the question (\textit{incorrect and unrelated}).

\paragraph*{Evaluation dimensions}
\label{sec:eval_dims}
\bench{} is built from a diverse set of musical works and their detailed descriptions, and serves as a foundation for evaluating Audio LLMs across various dimensions of music comprehension.
To delineate the specific evaluation dimensions encompassed by our benchmark, we develop a taxonomy consisting of two primary categories: \textit{knowledge} and \textit{reasoning}.\footnote{We include the full taxonomy in the online supplementary materials.}
Each category is further divided into several dimensions, informed by insights from national music education programs and existing research on music folksonomies  \cite{sordo_inferring_2013}.
This structured approach allows us to assess the depth and breadth of music-related competencies systematically, offering a holistic view of models' capabilities in the music domain.

\begin{figure}[t]
\centering
 \includegraphics[width=0.8\columnwidth]{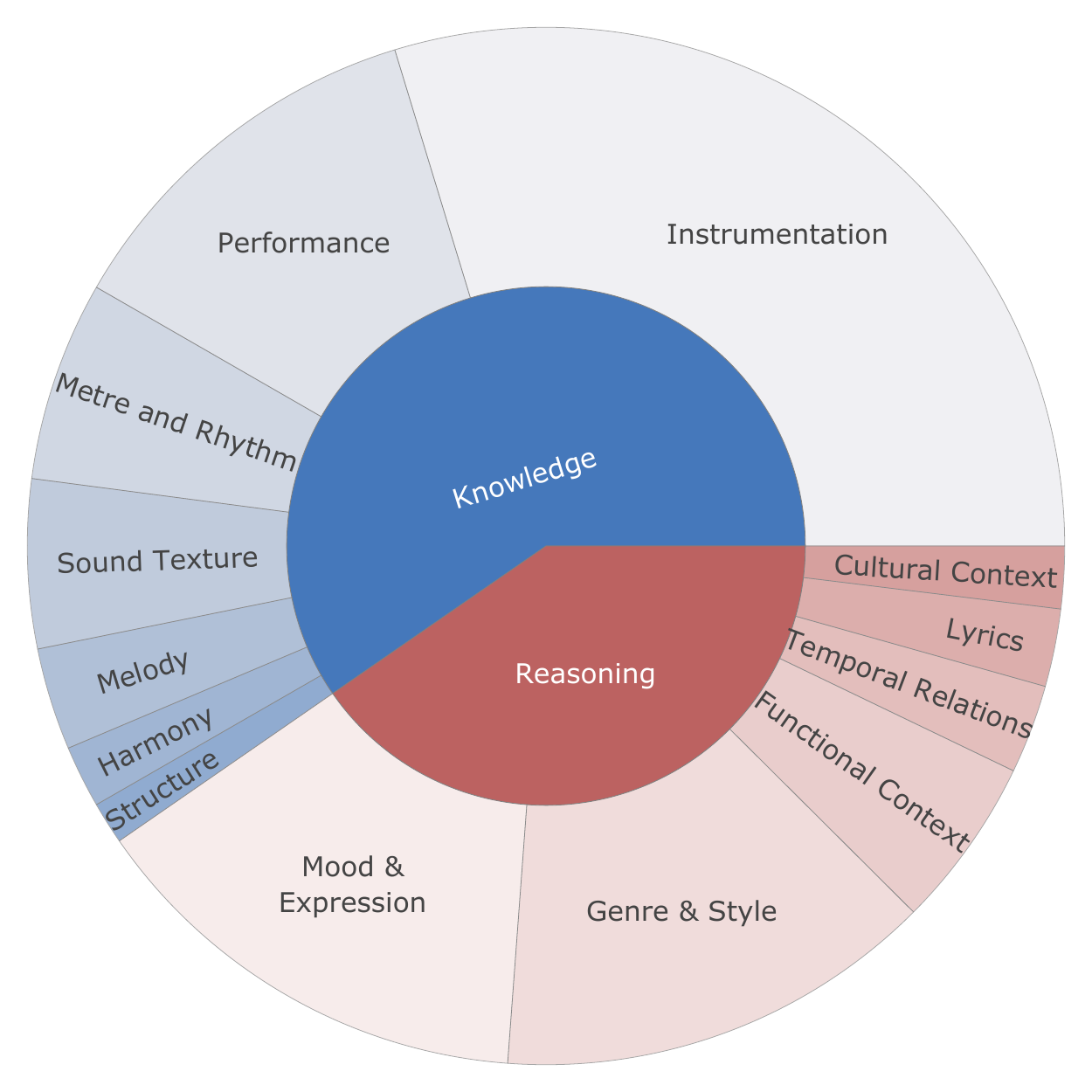}
   \vspace{-1em}
 \caption{\textbf{Distribution of evaluation dimensions} covered by MuChoMusic across knowledge and reasoning.}
 \label{fig:categories}
  \vspace{-1em}
\end{figure}

In the \textit{knowledge} category, questions probe a model's ability to recognise pre-acquired knowledge across various musical aspects
\begin{enumerate*}[label=(\roman*), before=\unskip{: }, itemjoin={{, }}, itemjoin*={{, and }}, after=.]
    \item melody
    \item harmony
    \item metre and rhythm
    \item instrumentation
    \item sound texture
    \item performance
    \item structure
\end{enumerate*}
Questions that test \textit{reasoning} are instead designed to require the synthesis and analytical processing of multiple musical concepts
\begin{enumerate*}[label=(\roman*), before=\unskip{: }, itemjoin={{, }}, itemjoin*={{, and }}, after=.]
    \item mood and expression
    \item temporal relations between elements
    \item interpretation of lyrics
    \item genre and style
    \item historical and cultural context
    \item functional context
\end{enumerate*}
An example of reasoning might involve using an understanding of tempo, chord quality, and instrumentation in concert to ascertain the mood of a music piece.
Each question can cover multiple dimensions and their categorisation is obtained automatically, as described in Section \ref{sec:data_curation}.
Figure \ref{fig:categories} shows the coverage of the two categories and their respective dimensions within the benchmark.
Over half the questions test at least one aspect of musical knowledge, such as features relating to instrumentation or performance characteristics, while 44\% are dedicated to probing reasoning skills.
While the distribution of dimensions within each category is not balanced, we note that this reflects the distribution of different musical concepts within music captions \cite{manco_song_2023}, resulting in categories such as instrumentation, mood and genre appearing more frequently.

\subsection{Dataset construction}
\label{sec:data_curation}
To build our dataset, we automatically transform human-written music captions into multiple-choice questions. These are then carefully validated by multiple human annotators, alongside the associated audio, in order to filter out invalid, ambiguous or irrelevant questions resulting from inaccuracies or hallucinations in the model output.

\paragraph*{Data sources}
We source our data from music caption datasets as we aim for elaborate and linguistically diverse information about the music.
Currently, only two captioning datasets provide sufficiently detailed music descriptions, namely the Song Describer Dataset (SDD) and MusicCaps.
SDD contains 2-minute-long music clips with single-sentence captions crowd-sourced from music enthusiasts, while the captions in MusicCaps, describing 10-second audio snippets, are written by professional musicians.
From SDD, we select all tracks that have at least two captions, to ensure enough information is provided to the model to be able to formulate interesting and challenging questions.
While this is not possible for the MusicCaps dataset, where only one caption is available for each track, we note that descriptions are, on average, longer than in SDD and designed to be more comprehensive. 
From the genre-balanced subset of the MusicCaps test split, we exclude all tracks for which the labels indicate a low recording quality, to prevent differences in audio quality from affecting the results.
For both datasets, we employ a state-of-the-art genre tagging model \cite{alonso2022music} to identify non-musical tracks and to sub-sample songs from the most common genres (e.g. rock and electronic). 
Through this curation process, we select 227 unique tracks from SDD and 497 from MusicCaps.
We supplement the descriptions with short text labels taken from the dataset itself in the case of MusicCaps and from a related dataset for SDD \cite{Bogdanov2019}.

\begin{figure}[t]
\centering
    \includegraphics[width=0.87\columnwidth]{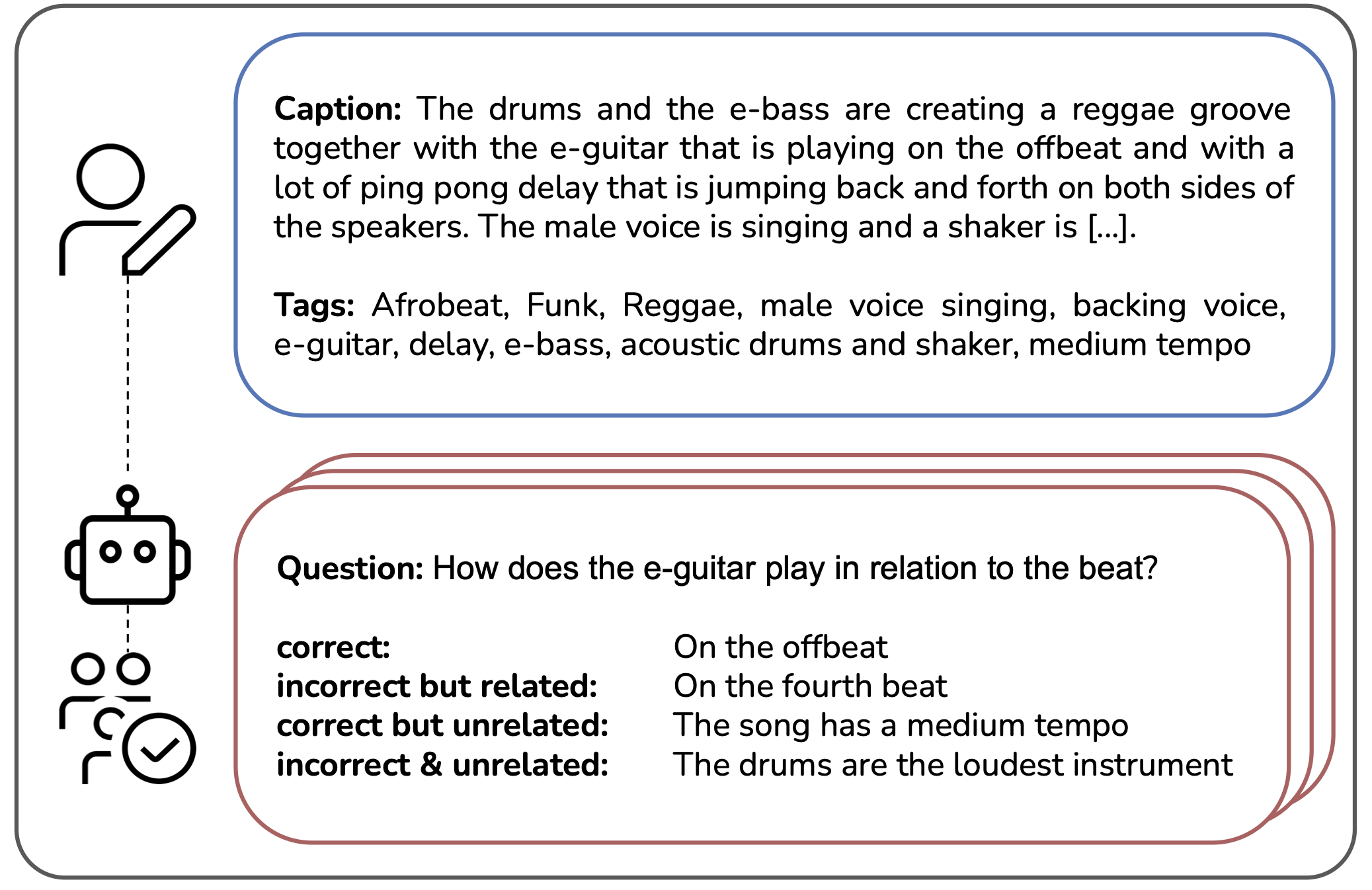}
   \vspace{-0.8em}
 \caption{\textbf{QA generation and validation pipeline}. Example shown here is from MusicCaps \cite{agostinelli_musiclm_2023}.}
 \label{fig:generation_example}
  \vspace{-0.5em}
\end{figure}

\paragraph*{QA generation}
We generate the question-answer sets by instructing Gemini 1.0 Pro \cite{gemini_team_gemini_2023} to formulate question and answer options for a given human-written caption. 
To leverage the model's in-context learning capability, we prompt it with a detailed task description and three examples of input (description and tags) and expected output.
In addition to the question and answer pairs, we ask the model to start its output with a summary of the provided information about the music recording and to interleave the distractor answer options with explanations of their suitability.
This way of prompting is inspired by the chain-of-thought methodology and helps to elicit the best model responses \cite{NEURIPS2022_8bb0d291,wei_cot-prompting_2024}.
This way, we obtain three multiple choice questions from each description on average and collect a total of 2,091 question-answer pairs.
An example of the generated questions is shown in Figure \ref{fig:generation_example}. 

\paragraph*{Data validation}
In order to ensure that questions and answers in our benchmark are factually accurate, aptly written and that each question can be correctly answered based on the available audio, we validate all sets of questions via human annotators. For this step, we recruit 222 participants via the Prolific platform (www.prolific.com).
During annotation, a question, the corresponding audio clip, and all four answer options are presented to the participants in random order, for a total of 30 to 50 question items.
Participants are then asked to select all options that correctly answer the question or skip the question by indicating that they are unable to provide an answer or that the question is not valid.
Following this procedure, for each question, we collect three to five annotations, stopping early if different annotators are in agreement.
This task setup is intended to vet questions and detect those that do not adhere to the intended multiple-choice format, either because the expected correct answer is not the only plausible option or because any one of the distractors is more likely.
Consequently, we exclude questions from our final dataset for which i) less than 50\% of the annotations indicate the intended correct answer or ii) more than 50\% of the annotations mark any of the disctractors as a plausible answer.
The final dataset comprises 858 questions from MusicCaps descriptions and the remaining 329 from SDD captions.

\paragraph*{Question categorisation}
Once questions are validated, we categorise them according to our taxonomy outlined in Section \ref{sec:eval_dims}. To achieve this, we employ Gemini 1.0 Pro, this time prompting it to automatically label each question with one or more of the evaluation dimensions.
The prompt includes the full taxonomy including detailed descriptions of all dimensions, a chain-of-thought instruction, and a single question with only the correct answer.
The produced output contains an explanation of the categories and dimensions assigned to each question.

\section{Benchmarking with MuChoMusic}
We now demonstrate the use of our benchmark, describing our proposed evaluation protocol and metrics, and then detailing our experiments on benchmarking Audio LLMs. 

\subsection{Evaluation Protocol}\label{sec:experimental_setup}
In multiple-choice-based evaluation, a model is provided with a question and a set of answer options, and is then tasked with selecting the most suitable answer. In practice, this can be accomplished in different ways \cite{liang2023holistic}. In our experiments, we adopt \textit{output-based} evaluation: given a music clip and an associated question-answer set, the language output produced by the model is mapped to one of the candidate options by string matching. 
Another common approach in MC evaluation is to determine the selected answer through the conditional log likelihood scores of the tokens forming each of the different options. While this can help estimate uncertainty and confidence in the model predictions, in our experiments, we explore only the output-based setting, for three reasons: (1) this corresponds to real-world use of the models, as interactions usually take the form of a conversation; (2) it has a lower computational cost; (3) prior work has demonstrated that sentence probabilities are not necessarily indicative of the probabilities assigned to the answers \cite{robinson_leveraging_2023}.
To extract the selected answer from the generated outputs, we match either the option identifier (\textit{A}, \textit{B}, \textit{C} or \textit{D}) or the full answer text, if one and only one is given in the output. 

\paragraph*{Metrics}
We look at two main metrics to measure model performance on our benchmark: accuracy and  instruction following rate (IFR). Accuracy is given by the percentage of correctly answered questions out of the total set of questions. IFR is given by the percentage of generated answers that correspond to one of the given options. In both cases, finegrained scores can be obtained by considering only the subset of questions covering at least one of the available evaluation dimensions shown in Figure \ref{fig:categories}.

\paragraph*{Adaptation}
An important design factor in the evaluation of LLM-based models is adaptation \cite{liang2023holistic}, the process of adapting the input to a suitable format. While the format of the audio input is typically fixed by the model design, text inputs allow for more flexibility and different prompting techniques have been shown to significantly influence model's behaviour \cite{wei_cot-prompting_2024,NEURIPS2022_8bb0d291,DBLP:conf/iclr/WeiBZGYLDDL22}.
Beyond simply passing the question and answer options as the input text, corresponding to \textit{zero-shot prompting}, an effective alternative strategy is to leverage \textit{few-shot in-context learning} (ICL), whereby the model is presented with a set of reference inputs that exemplify the task prior to being shown the question of interest.
We experiment with in-context learning in our experiments, providing between 0 and 5 examples in the text input. In the interest of standardisation and to ensure a fair comparison between the models, unless otherwise specified, we keep the prompt selection fixed in our final experiments, following an initial exploration.

\begin{table}
\small
    \centering
    \begin{tabular}{p{2.25cm}>{\centering\arraybackslash}m{2.4cm}>{\centering\arraybackslash}m{2.35cm}}
        \toprule
         Model & Audio encoder & LLM  \\
         \midrule
        MusiLingo \cite{deng_musilingo_2023} & MERT \cite{li2024mert}  & Vicuna
7B \cite{vicuna2023} \\
        MuLLaMa \cite{liu_music_2024}  & MERT \cite{li2024mert} & LLaMA-2 7B \cite{touvron_llama_2023}    \\
        M2UGen \cite{hussain_m2ugen_2024}  & MERT \cite{li2024mert} & LLaMA-2 7B \cite{touvron_llama_2023} \\
        \cdashlinelr{1-3}
        SALMONN \cite{tang_salmonn_2024} &  BEATS \cite{chen_beats_2023} \& Whisper\textsubscript{large-v2} \cite{radford_robust_2023} & Vicuna 7B \cite{vicuna2023}\\
        Qwen-Audio \cite{chu_qwen-audio_2023} & Whisper\textsubscript{large-v2} \cite{radford_robust_2023} & Qwen 7B \cite{qwen2023} \\
        \bottomrule
    \end{tabular}
    \caption{\textbf{Overview of models} we evaluate in our study.}
    \label{tab:baselines}
\end{table}

\subsection{Models}
In our evaluation, we consider three music-specific models, MuLLaMA \cite{liu_music_2024}, MusiLingo \cite{deng_musilingo_2023}, and M2UGen \cite{hussain_m2ugen_2024}, and two general-audio LLMs which can be applied to music, as reported in their respective papers, SALMONN \cite{tang_salmonn_2024} and Qwen-Audio \cite{chu_qwen-audio_2023}.
To the best of our knowledge, these are all the existing Audio LLMs which can be applied to music and for which open-source weights are available.
These all share a similar architectural design and are composed of a backbone LLM, an audio encoder and a lightweight learnable adapter module to align embeddings produced by the audio encoder to the input space of the LLM, based on either the LLaMA-adapter \cite{zhang_llama-adapter_2024} (MuLLaMA, MusiLingo, M2UGen) or a Q-Former network \cite{li_blip-2_2023} (SALMONN). An overview of the backbones used in each model is provided in Table \ref{tab:baselines}.
All systems are trained via instruction tuning \cite{wang-etal-2023-self-instruct,DBLP:conf/iclr/WeiBZGYLDDL22} and all employ a combination of different instruction datasets, often in multiple training stages including pre-training and fine-tuning.
For all models, we follow the official implementation and use default inference settings.
We repeat all experiments 3 times, randomly shuffling the order in which answer options are presented, and report average performance across all runs.

\section{Results and Discussion}\label{sec:results}
In this section, we first presents findings from our benchmarking experiments, with the goal of elucidating the current state of music understanding in Audio LLMs. We then illustrate how \bench{} can be used to derive new insights via a diagnostic analysis, and discuss key takeaways.

\subsection{Benchmarking Results}
\begin{figure}[t]
\centering
 \includegraphics[scale=0.43]{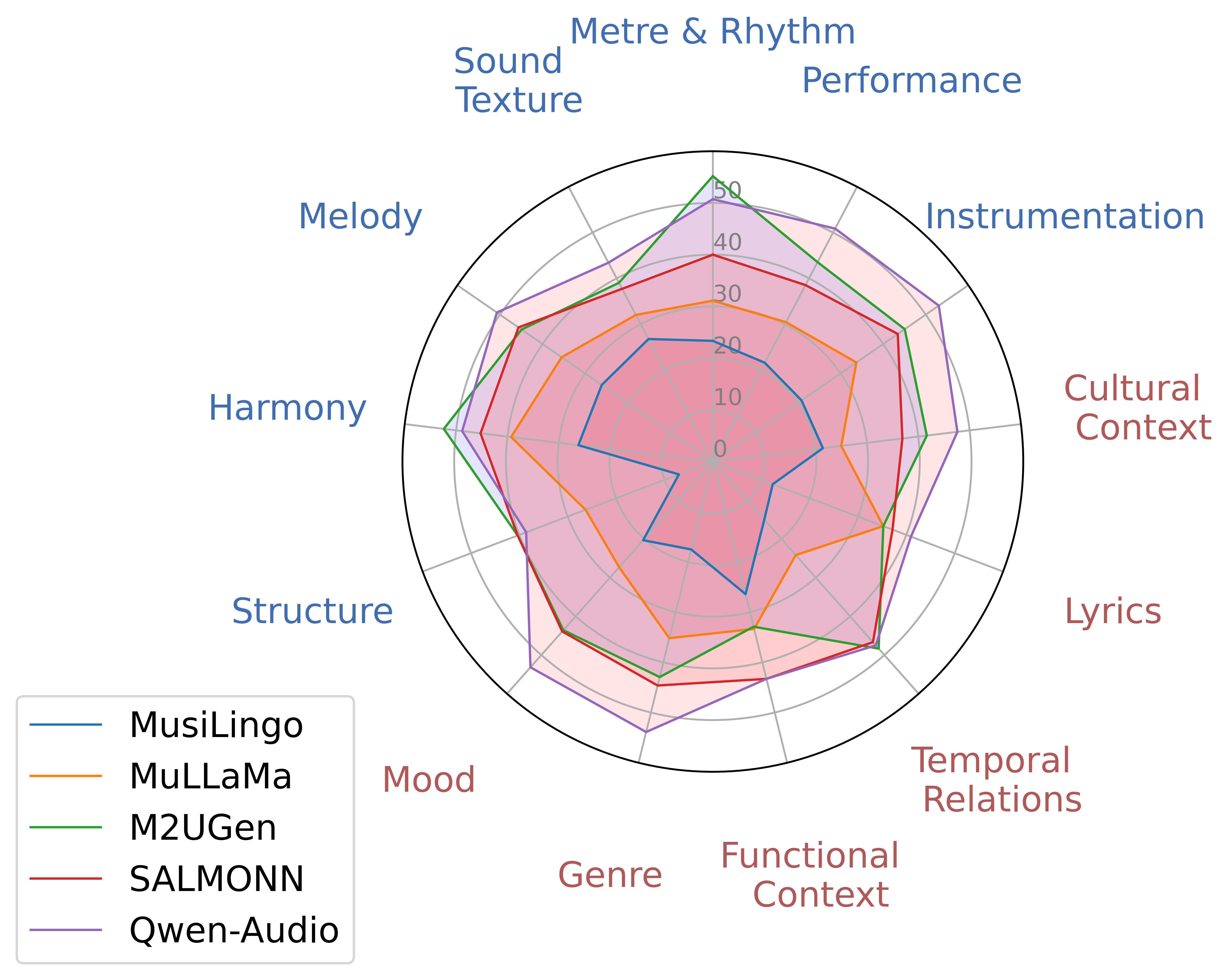}
 \caption{\textbf{Finegrained accuracy} across evaluation dimensions in knowledge (labelled in blue) and reasoning (red).}
 \label{fig:eval_dims}
\end{figure}

We report results for all models in Table \ref{tab:results}, showing the overall accuracy score alongside detailed scores on knowledge and reasoning questions, and the instruction following rate (IFR). Figure \ref{fig:eval_dims} presents a breakdown of accuracy scores along all reasoning and knowledge dimensions. Unless otherwise specified, we show one-shot performance for all models, as we find this to be the overall optimal setting, as we discuss in more detail in Section \ref{sec:diagnosis}. From this, we observe that current models generally perform poorly across all settings and along all evaluation dimensions. Among these, Qwen-Audio stands out with a score of 51.4\%.
Surprisingly, with the exception of M2UGen, music-specialised models generally perform worse than general-audio ones, in some cases performing only marginally above or even below random performance.
As evidenced by the IFR, these models struggle to output answers in the correct format, which in turn negatively impacts their accuracy score. As shown later in Section \ref{sec:case_study}, we find that, when none of the answer options is selected by the model, this is often due to \textit{auditory hallucinations}, \textit{language hallucinations} or \textit{training biases}. 

\subsection{Analysis and Discussion}\label{sec:diagnosis}
We now investigate factors influencing performance along different axes by using our benchmark as a diagnostic tool.

\begin{table}[t]
    \centering
    \small
    \begin{tabular}{lc>{\centering\arraybackslash}m{1.2cm}>{\centering\arraybackslash}m{1.2cm}c}
    \toprule
        \multirow{2}{*}{Model}& \multicolumn{3}{c}{Accuracy} & IFR\\
        \cmidrule(lr){2-4} \cmidrule(lr){5-5}
         & All & Knowledge  & Reasoning  & All \\ 
        \midrule 
        MusiLingo \cite{deng_musilingo_2023} & 21.1 & 22.0 & 19.2 & 71.6 \\
        MuLLaMa \cite{liu_music_2024}  & 32.4 & 32.3 & 31.3 & 79.4 \\
        M2UGen \cite{hussain_m2ugen_2024}  & 42.9 & 44.9 & 41.2 & 96.4 \\
        \cdashlinelr{1-5}
        SALMONN \cite{tang_salmonn_2024} & 41.8 & 41.0 & 43.3 & 99.8 \\
        Qwen-Audio \cite{chu_qwen-audio_2023} & 51.4 & 51.1 & 51.0 & 89.7 \\
        \cdashlinelr{1-5}
        \textit{Random guessing} & 25.0 & 25.0 & 25.0 & 100.0 \\
        \bottomrule
    \end{tabular}
        \caption{Overall \textbf{benchmarking results}.}
    \label{tab:results}
     \vspace{-1em}
\end{table}

\begin{figure}[t]
\centering
 \includegraphics[width=\columnwidth]{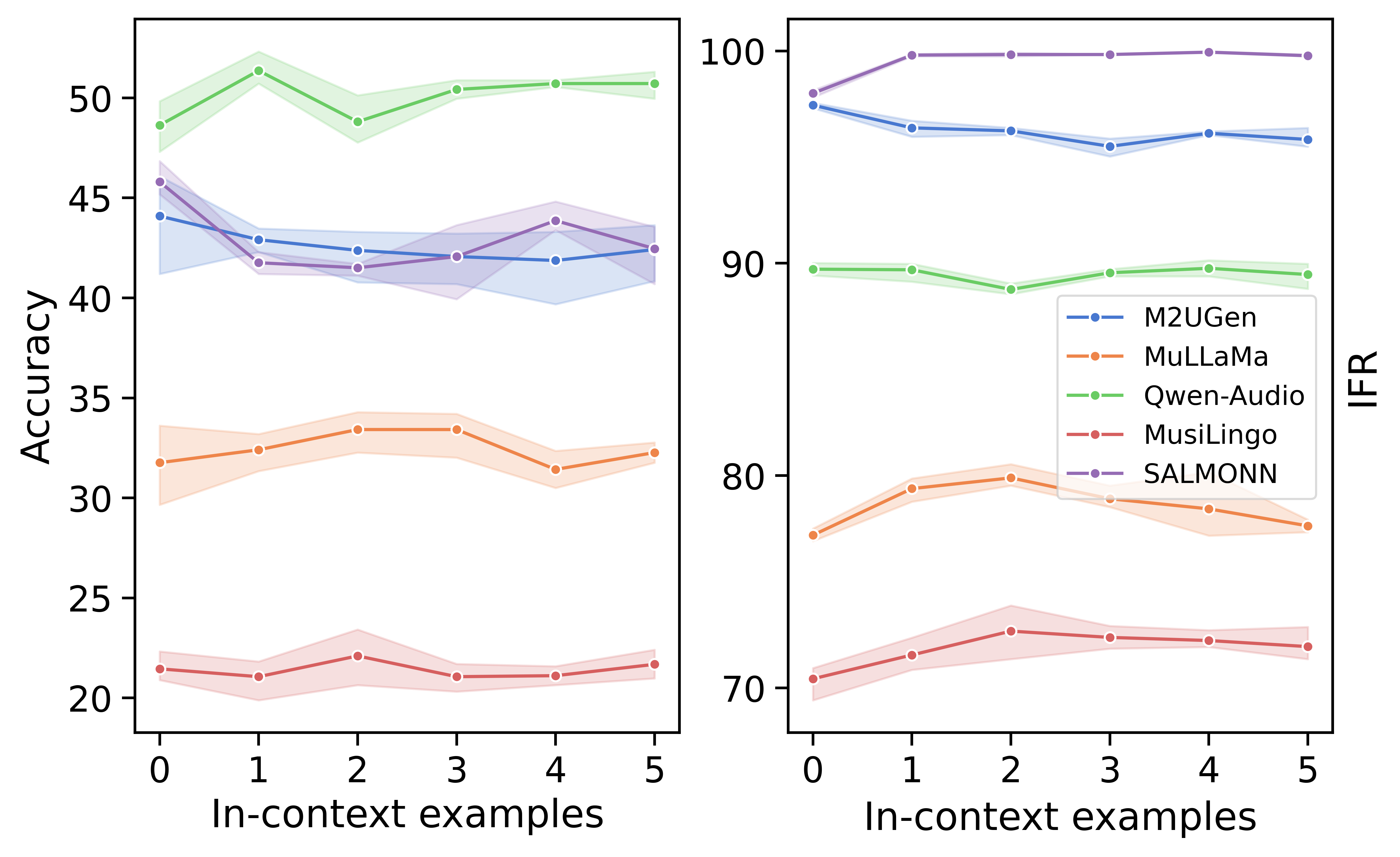}
 \caption{\textbf{Effect of the number of in-context examples} on accuracy (left) and instruction-following rate (right).} 
 \label{fig:acc_vs_shots}
\end{figure}

\paragraph*{Are models sensitive to prompts?} We first study the effect of varying the number of in-context examples. As shown in Figure \ref{fig:acc_vs_shots}, providing a single example is occasionally beneficial to accuracy and IFR, but with both the difference magnitude and overall impact differing between models. Additionally, this trend does not hold after the one-shot setting, and we see no consistent improvement when using a larger number of examples. Interestingly, we observe that, for M2UGen, Qwen-Audio and MuLLaMa, changes in accuracy from zero- to one-shot prompts are accompanied by a reduction in variance, suggesting that ICL can help minimise variability in the model output. While we do not explore this in our experiments, we also hypothesise that the advantages of ICL may become more prominent through multimodal few-shot prompting \cite{doveh_towards_2024, zhao_mmicl_2024}, which we leave for future work.

\begin{figure}[t]
\centering
 \includegraphics[width=\columnwidth]{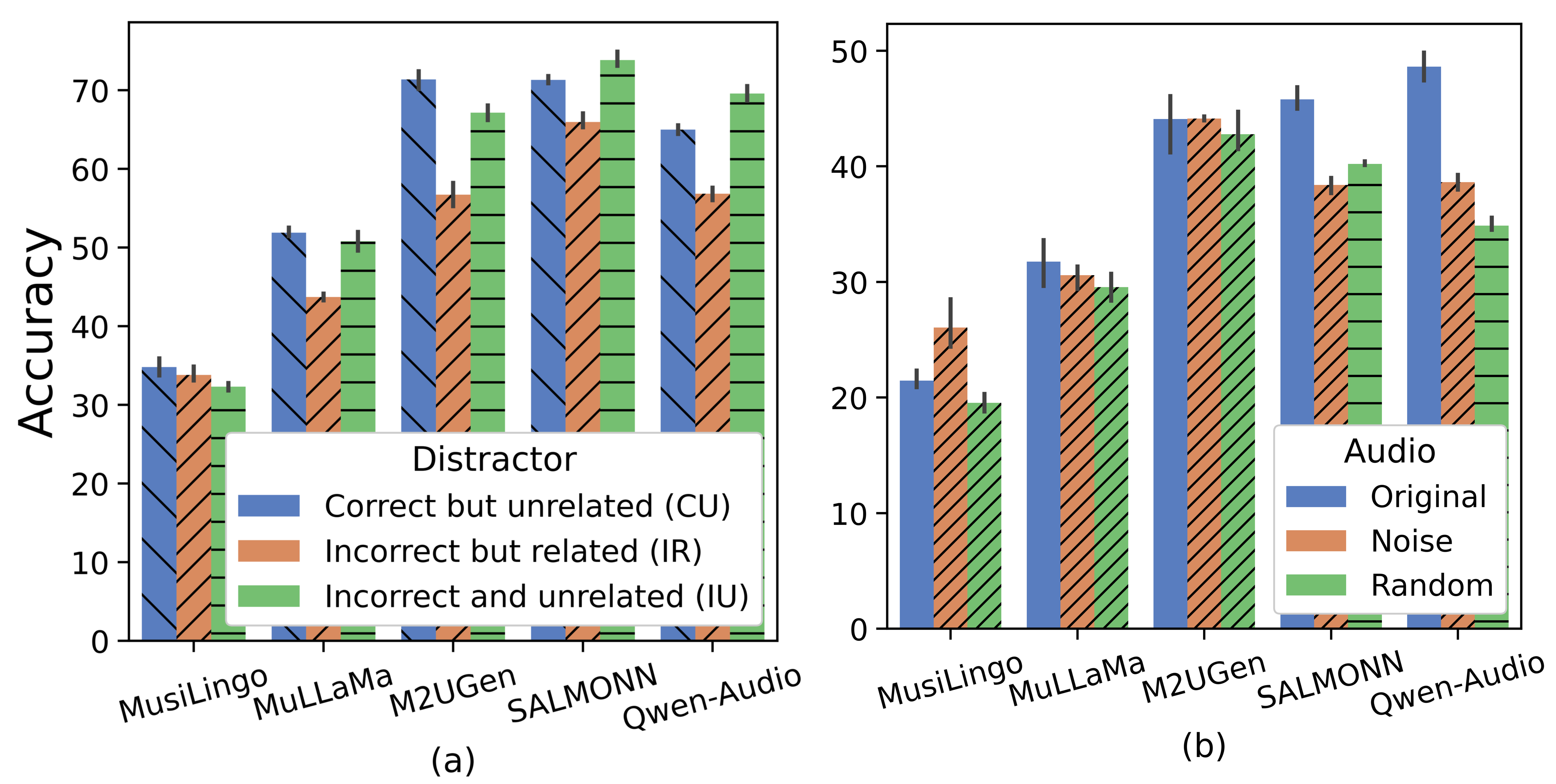}
 \caption{(a) \textbf{Effect of using different types of distractors}: models tend to perform worst when tasked with distinguishing between two related answers. (b) \textbf{Audio attention test}: only some models display a significant drop in performance when provided with incorrect audio inputs. For these experiments, we adopt zero-shot prompting.}
 \label{fig:audio}
\end{figure}

\paragraph*{How do models respond to different distractors?}
Next, we shift our attention to examining how distractors in our benchmark influence the difficulty of the task. To this end, we ablate answer options corresponding to the different kinds of distractors described in Section \ref{sec:data_curation}, and present the model with only two or three answer options. In Figure \ref{fig:audio}(a) we show how performance is affected when using only one distractor alongside the correct option, always randomising their order. From this, we observe that the two distractors containing information which is not related to the question (CU and IU) have a similar effect, while including the \textit{incorrect but related} (IR) option consistently makes the task more challenging. This phenomenon persists when adding a second distractor (not shown here), with combinations which include IR invariably leading to worse performance. Intuitively, the two \textit{unrelated} options can be ruled out based on the text input only, while selecting the correct answer between two options that appear relevant requires engaging multimodal understanding to relate information in the audio content to the text in the question. Crucially, this indicates that models particularly struggle to discern between options that are equally plausible based on the text input only, suggesting that less attention is given to the audio content. This forms the basis of our hypothesis that current Audio LLMs are characterised by a strong language bias, leading to poor performance in tasks that are more audio-dependent. We test this hypothesis in the next section.

\paragraph*{Do models actually pay attention to the audio?} In order to verify whether the audio input is effectively being ignored or is overshadowed by its text counterpart, we devise a simple test, which we call \textit{audio attention test}, where we replace the audio clip corresponding to a given question with either white Gaussian noise or a randomly chosen track from the dataset. In order to pass this test, a model should display a statistically significant drop in performance when either form of audio perturbation is used, compared to its baseline performance. We showcase results on this test in Figure \ref{fig:audio}(b). From this, we clearly see that, with the exception of SALMONN and Qwen-Audio, all models fail the audio attention test, and the severity of this failure is often negatively correlated to their overall performance on the benchmark (see Table \ref{tab:results}). This confirms that current Audio LLMs are biased towards textual information, often choosing answers that score well under their language prior. Additionally, it provides an explanation for their low performance on the benchmark, as this is effectively bounded by the maximum score they can attain mostly based on the language input. We argue that this constitutes a major pitfall in the design and training procedure of these models, which results in music understanding abilities that do not match the expected performance, as obtained through prior evaluations. 

\subsection{Failure Modes}\label{sec:case_study}
While the core goal of our benchmark is to provide standardised automatic evaluation to objectively measure general music understanding capabilities, we argue that it can also constitute a useful tool for qualitative assessment.
We showcase three examples here, focusing on the two lowest-performing models.
While this is not an exhaustive analysis, these examples offer a bird's-eye view of how language pre-training biases percolate through multimodal training, resulting in failures to attend to the inputs in our evaluation.
To describe these, we borrow terminology from \cite{guan_hallusionbench_2024}.

\paragraph*{Auditory hallucination} One of the ways models fail to provide a suitable answer falls under the category of \textit{auditory hallucination}, whereby a response includes references to musical elements that are not present in the audio. For example, when asked about an accompaniment instrument, models with this type of hallucination may ignore any suitable option provided (\textit{``acoustic guitar''} or \textit{``strings''}), instead answering \textit{``The song is accompanied by a piano.''}, when the audio clip clearly contains no piano.

\paragraph*{Language hallucination} Another instance of hallucination concerns mundane statements that deviate from the topic of the question altogether.
Among others, an observed case of this failure mode is a statement of the form \textit{``The song has a clear and coherent rhythm structure''} to a question specifically asking about the \textit{``type of drum beat''}.

\paragraph*{Training data bias}
The last failure mode we encounter is related to a bias towards frequent patterns occurring in the training data.
While some of the benchmarked models undergo a stage of training that includes instruction-tuning examples with questions and answers, occasionally they still produce trivial outputs.
For example, when asked \textit{``What is the intended purpose of this song?''}, a model with this type of bias may answer \textit{``The intended purpose of this song is not mentioned in the caption''}. 
Reviewing MusicQA, used in training MuLLaMa and MusiLingo, reveals that a high number of the LLM-generated training examples mention similar phrases, thus likely biasing the model towards this type of uninformative but highly likely output.

\section{Conclusion}\label{sec:conclusion}
We have presented \bench{}, a multiple-choice music question answering benchmark designed to test music understanding in Audio LLMs.
From an evaluation of five state-of-the-art systems, we find that our benchmark acts as a challenging and informative test, and that current models do not yet leverage both the audio and text modalities fully.
All questions in our benchmark are synthesised from human-written music descriptions and manually reviewed to guarantee  high data quality. 
A categorisation of the questions highlights that \bench{} offers a broad coverage of areas targeted by current models, and additionally pinpoints gaps that could guide future developments in the field. 
While we demonstrate that our approach leads to new insights, we note that the multiple-choice format presents some limitations \cite{zheng_large_2024}. Therefore evaluation on \bench{} should be complemented via further benchmarking efforts to address additional aspects of music understanding through different tasks and formats. 

\section{Ethics Statement}
\subsection{Annotator welfare}
Prior to participation, the annotation experiment described in Section \ref{sec:data_curation} was approved by the Queen Mary Ethics of Research Committee to ensure alignment with ethical guidelines and protections for human subjects in research. We did not collect any personal data from our annotators, safeguarding their privacy and confidentiality. Annotators were fully informed about the objectives of the research, the nature of their tasks, and the use of their annotations, underpinning their informed consent before contributing to the project.
In an effort to provide a fair compensation for their contributions, annotators were paid £9 per hour.

\subsection{Biases and fairness}
In constructing the \bench{} benchmark, our data collection strategy included sourcing music tracks from a variety of backgrounds, acknowledging the inherent challenges in representing the rich diversity of global music cultures within our dataset. We recognise that our initiative does not fully balance the benchmark across all genres, languages, and cultural backgrounds, and annotations were conducted exclusively in English due to logistical constraints, highlighting areas for future expansion and improvement.

\section{Acknowledgements}
IM is a research student at the UKRI Centre for Doctoral Training in Artificial Intelligence and Music, supported jointly by UK Research and Innovation [grant number EP/S022694/1] and Universal Music Group. EB is supported by RAEng/Leverhulme Trust research fellowship LTRF2223-19-106.

\printbibliography[title=References, heading=bibnumbered]

\end{document}